\documentclass[aps,prd,superscriptaddress,preprintnumbers,tightenlines,nofootinbib, twocolumn,10pt,floatfix,amsmath,amssymb,longbibliography,superscriptaddress,nofootinbib]{revtex4-2}

\usepackage{graphicx}
\usepackage{dcolumn}
\usepackage{bm}
\usepackage{tabularx}
\usepackage{multirow}
\usepackage{capt-of}
\usepackage{color}
\usepackage{url}
\usepackage[utf8]{inputenc}
\usepackage{xcolor} 
\usepackage[table]{xcolor}
\usepackage[dvipsnames]{xcolor}
\usepackage[svgnames]{xcolor}
\usepackage[x11names]{xcolor}
\usepackage{booktabs}

\usepackage[nolist,nohyperlinks]{acronym}
\usepackage{xspace}
\usepackage[colorlinks]{hyperref}
\usepackage[normalem]{ulem}
\usepackage{longtable}
\usepackage{makecell}

\definecolor{dodgerblue}{HTML}{1E90FF}
\definecolor{viennared}{HTML}{DA0A14}
\definecolor{ctorange}{HTML}{FF6C0C}
\definecolor{granadagreen}{HTML}{078931}
\definecolor{wales}{HTML}{ff0038}
\definecolor{valenciacfred}{HTML}{ee3524}
\definecolor{barcelonafcgold}{HTML}{edbb00}
\definecolor{jam}{HTML}{A50B5E}
\definecolor{austriawien}{HTML}{441678}

\AtBeginDocument{%
  \hypersetup{
    citecolor=dodgerblue,
    linkcolor=dodgerblue,   
    urlcolor=dodgerblue}}

\usepackage{amsmath}
\usepackage{cancel}
\usepackage{amsfonts}
\usepackage{amssymb}
\usepackage{bm}
\usepackage[colorlinks]{hyperref}
\usepackage{mathrsfs}
\usepackage{graphicx}
\usepackage{multirow}
\usepackage{empheq}
\usepackage{ulem}
\usepackage{tensor}
\usepackage{tabularx}
\usepackage{cleveref}
\usepackage{stmaryrd}
\newcolumntype{Y}{>{\centering\arraybackslash}X}

\allowdisplaybreaks

\DeclareSymbolFontAlphabet{\mathrsfs}{rsfs}
\DeclareMathAlphabet{\mathcal}{OMS}{cmsy}{m}{n}

\usepackage{parskip}
\definecolor{darkgreen}{rgb}{0,0.5,0}

\hypersetup{
    unicode=false,          
    pdftoolbar=true,        
    pdfmenubar=true,        
    pdffitwindow=false,     
    pdfstartview={FitH},    
    pdftitle={My title},    
    pdfauthor={Author},     
    pdfsubject={Subject},   
    pdfcreator={Creator},   
    pdfproducer={Producer}, 
    pdfkeywords={keyword1} {key2} {key3}, 
    pdfnewwindow=true,      
    colorlinks=true,       
    linkcolor=red,          
    citecolor=cyan,        
    filecolor=magenta,      
    urlcolor=darkgreen,           
    linktocpage=true}

\newcommand{\be}{\begin{equation}}
\newcommand{\ee}{\end{equation}}

\newcommand{\sigmatp}{\widetilde{\sigma}^{(2)}_{+}}
\newcommand{\sigmatm}{\widetilde{\sigma}^{(2)}_{-}}

\newcommand\calO{{\mathcal{O}}}
\newcommand{\dd}{\mathrm{d}}

\newcommand{\nn}{\nonumber}
\newcommand{\fid}{\dot{\phi}}
\newcommand{\rd}{\dot{r}}
\newcommand{\Et}{\widetilde{E}}
\newcommand{\Lt}{\widetilde{L}}
\newcommand{\tmass}{m}

\newcommand{\Lag}{\mathcal{L}}
\newcommand{\de}{\text{e}}

\usepackage{ulem}
\allowdisplaybreaks

\usepackage{etoolbox}
\makeatletter
\patchcmd{\frontmatter@abstract@produce}
  {\vskip200\p@\@plus1fil
   \penalty-200\relax
   \vskip-200\p@\@plus-1fil}
\makeatother

\begin{document}

\title{Periastron advance from the perturbed Binet equation}

\author{Quentin \textsc{Henry}}
\email{quentin.henry@uib.es}
\affiliation{Departament de Física, Universitat de les Illes Balears, IAC3 – IEEC, Crta. Valldemossa km 7.5, E-07122 Palma, Spain}

\date{\today}

\begin{abstract}
We apply the Poincaré–Lindstedt method to the perturbed Binet equation describing conservative dynamics of the perturbed Kepler problem. We show that a simple treatment of divergent terms directly determines the frequency shift of the radial motion and hence the periastron advance. We provide explicit expressions for polynomial, logarithmic and inverse power perturbations, extend the procedure to higher perturbative orders and to problems involving several perturbation parameters. Applying this method to different physical effects relevant to compact binaries, we recover known results. We also obtain new contributions to the periastron advance, including the leading-order effects of arbitrary mass- and current-type tidal multipoles and mass-type spin-induced multipoles, the NNLO current-type tidal quadrupole contribution and electromagnetic electric-dipole contributions to NNLO. Finally, we illustrate how the method can be directly applied to the problem of particle motion in Reissner–Nordström, de Sitter-Schwarzschild metrics and in a Yukawa potential.
\end{abstract}


\maketitle

\section{Introduction}\label{sec:intro}

Perturbative solutions of periodic differential equations can contain secular terms, i.e. contributions that grow with the independent variable even though the exact motion remains bounded. Such terms are not physical features of the motion, but artifacts of a perturbative expansion performed around an unperturbed frequency. Their treatment has played an important role in celestial mechanics since the nineteenth century as Poincaré wrote in Les Méthodes Nouvelles de la Mécanique Céleste~\cite{poincare1892methodes}, “\textit{Aussi tous les efforts des géomètres, dans la seconde partie de ce siècle, ont-ils eu pour but principal de faire disparaître les termes séculaires. La première tentative sérieuse qui ait été faite dans ce sens est celle de Delaunay, dont la méthode est encore appelée sans doute à rendre bien des services.}”\footnote{Accordingly, all the efforts of mathematicians during the second half of this century have had as their principal aim the elimination of the secular terms. The first serious attempt in this direction was that of Delaunay, whose method is doubtless still destined to render many valuable services.}

Several perturbative techniques were developed to handle secular terms, including canonical perturbation theory and the Delaunay method~\cite{Delaunay:1860}, Hill-type methods~\cite{Hill}, averaging and the Krylov–Bogoliubov–Mitropolsky method~\cite{krylov1950introduction,Bogoliubov:1961}, multiple-scale methods~\cite{Nayfeh:1973}, and the Lindstedt method~\cite{Lindstedt}, later refined by Poincaré~\cite{poincare1892methodes}. In this work we use the latter and apply it to a perturbed Kepler problem written in terms of the Binet equation.

For a perturbed Binet equation, a naive perturbative solution generally generates a resonant harmonic and hence a secular term. The Poincaré–Lindstedt method introduces a frequency parameter which is expanded perturbatively and fixed by requiring the cancellation of this resonant contribution. In the Kepler problem, this frequency shift directly determines the periastron advance. At linear order in the perturbation, only the first Fourier coefficient of the perturbing function evaluated on the Keplerian orbit is needed. For polynomial perturbations this coefficient can be obtained algebraically, so that no integration is required. We also derive explicit formulas for logarithmic perturbations and inverse powers, and extend the construction to higher perturbative orders and to systems involving several perturbation parameters.

We then apply the method to a number of conservative orbital problems. For post-Newtonian compact binaries, the perturbed Binet equation can be obtained straightforwardly once the conserved energy and angular momentum are known. We recover several known results for point-mass, tidal, spin, and electromagnetic effects, and derive new contributions for higher multipoles and electromagnetic interactions. The method can also be applied directly whenever the orbital dynamics is already available in Binet form, as illustrated with particle motion in Reissner–Nordström and de Sitter-Schwarzschild spacetimes. Finally, we show that logarithmic terms appearing in standard harmonic coordinates can be treated directly without first transforming them away when one is interested only in the periastron advance.

The paper is organized as follows. In Sec.~\ref{sec:method} we present the Poincaré–Lindstedt treatment of a generic perturbed Binet equation and derive the formulas used throughout the paper. In Sec.~\ref{sec:PN} we apply the method to post-Newtonian, tidal, spin, electromagnetic, and test-particle dynamics. We conclude in Sec.~\ref{sec:ccl}.  \newpage

\section{The Poincaré-Lindstedt method}\label{sec:method}

In this section, we present the Poincaré–Lindstedt method, see e.g.~\cite{mickens1996oscillations} for a pedagogical explanation. We first define secular terms and illustrate their origin with a simple example, before treating a generic perturbation at linear order. We then consider several classes of perturbing functions useful for the applications to the perturbed Kepler problem of Sec.~\ref{sec:PN} and extend the procedure to higher perturbative orders and to problems involving several small parameters.

\subsection{Illustrating the problem}\label{subsec:example}

Consider the simple differential equation
\begin{equation}
f''(x) + f(x) = \cos x\,,
\end{equation}
with initial conditions $f(0) = 1$ and $f'(0) = 0$. Its unique solution reads
\begin{equation}
f(x) = \cos x +\frac{x}{2} \sin x\,.
\end{equation}
The second term is divergent since it is unbounded as $x$ grows. Throughout this paper, we will refer to these such as secular terms. More generally, consider a periodic source function $F$ such that
\begin{equation}
f''(x)+ \Omega^2 f(x) = F(x)\,.
\end{equation}
Secular terms arise whenever $F$ contains a harmonic resonant with the homogeneous solution, i.e. proportional to $\cos(\Omega x)$ or $\sin(\Omega x)$. For example, suppose the source function $F = \sum_k \cos(k \omega x)$, then a secular term is generated if there exists a harmonic such that $k\omega=\Omega$.

Let us now illustrate how such terms arise in a perturbative problem. Suppose that $\varepsilon\ll 1$ and consider the simplest example
\begin{equation}\label{eq:PBEex}
y''(\phi) + y(\phi) = 1 + \varepsilon\,y^2(\phi)\,.
\end{equation}
A naive perturbative treatment consists in writing $y=y_0+\varepsilon y_1+\calO(\varepsilon^2)$ and solving the equation order by order. The solution at lowest order reads $y_0 = 1+ A \cos\Delta\phi$, with $\Delta\phi = \phi - \phi_0$. The first-order equation is then
\begin{equation}
y_1''(\phi) + y_1(\phi) = 1+\frac{A^2}{2}\left[1+\cos\bigl(2\Delta\phi\bigr)\right]+ 2 A \cos\Delta\phi \,.
\end{equation}
For $A\neq 0$, the last term is resonant with the homogeneous solution. It would therefore generate a secular term in the solution for $y_1$, which is problematic since the solution of~\eqref{eq:PBEex} is bounded. Its presence results from expanding the solution around the unperturbed frequency. The Poincaré–Lindstedt method removes this divergent behavior by allowing the frequency itself to receive perturbative corrections.

\subsection{Linear-order solution}\label{subsec:linorder}

Let us consider the differential equation
\begin{equation}\label{eq:yddgen}
y''+y = 1 + \varepsilon\,f(y) \,,
\end{equation}
where $y'= \tfrac{\dd y}{\dd\phi}$ and $f$ is assumed to be continuous and differentiable\footnote{This hypothesis can be relaxed, since at linear order it does not need to be differentiable. This method also works if the perturbing function contains the first derivative $y'$.} and not to depend explicitly on the independent variable $\phi$. To remove the secular terms described in Sec.~\ref{subsec:example}, the Poincaré–Lindstedt method introduces a new independent variable $\theta = \omega\, \phi$, where the frequency~$\omega$ is also treated perturbatively. At linear order,
\begin{subequations}\label{eq:yomegalin}
\begin{align}
y(\theta,\varepsilon) &= y_0(\theta) + \varepsilon\, y_1(\theta) + \calO(\varepsilon^2)\,,\\
\omega(\varepsilon) &= \omega_0 + \varepsilon\, \omega_1 + \calO(\varepsilon^2)\,.
\end{align}
\end{subequations}
The constant $\omega_0$ is chosen to match the frequency of the unperturbed equation which is 1 here. The leading order solution is $y_0(\theta) = 1 + e\cos\theta$, where $e$ is a constant that will represent the eccentricity in Sec.~\ref{sec:PN}. When inserting these expressions in~\eqref{eq:yddgen}, the linear order reads
\begin{equation}\label{eq:yddfy0}
\frac{\dd^2 y_1}{\dd\theta^2} + y_1 = 2 \omega_1( y_0 - 1 ) + f(y_0) \,.
\end{equation}
To exhibit the $\cos\theta$ and $\sin \theta$ harmonics of $f$ in~\eqref{eq:yddfy0}, we turn to the Fourier series of the perturbing function. Noting that $f(y_0)$ is the composed function $g= f\circ y_0$, i.e. $g(\theta) = f\bigl(y_0(\theta)\bigr)$, then $g$ is a $2\pi$-periodic, continuous and differentiable function (because so are $f$ and $y_0$). Thus, we can apply the Dirichlet theorem, which states that the function is equal to its Fourier series at each point of continuity. Furthermore, $g$ is an even function hence, when employing the real formulation of Fourier series, all the sine coefficients $b_n(g)$ vanish. Thus, $\forall \theta \in\mathbb{R}$,
\begin{equation}
g(\theta) = \sum_{k=0}^\infty a_k(g) \cos(k\theta)\,,
\end{equation}
where the cosine coefficients of $g$ are taken as
\begin{subequations}
\begin{align}
a_0(g) &=\frac{1}{\pi}\int_0^\pi g(t)\,\dd t\,,\\
a_{k>0}(g) &= \frac{2}{\pi}\int_0^\pi g(t) \cos(kt)\,\dd t\,.
\end{align}
\end{subequations}
This means that we can rewrite~\eqref{eq:yddfy0} as
\begin{equation}\label{eq:d2y1lin}
\frac{\dd^2 y_1}{\dd\theta^2} + y_1 = a_0(g) + \Bigl[ 2e\,\omega_1 + a_1(g)\Bigr] \cos\theta + \sum_{k=2}^\infty a_k(g)\cos (k\theta)\,.
\end{equation}
Secular terms will necessarily arise if $2e \,\omega_1+a_1(g)\neq 0$. Thus, by fixing $\omega_1 = - a_1(g)/2e$, they will not appear. Finally, after imposing this condition, the solutions of~\eqref{eq:d2y1lin} are given by
\begin{equation}\label{eq:y1}
y_1(\theta) = A\cos\theta + B\sin\theta 
+ \sum_{\substack{k\in \mathbb{N}\\k\neq 1}}\frac{a_k(g)}{1-k^2}\cos (k\theta)\,,
\end{equation}
where $A$ and $B$ are fixed by initial conditions\footnote{In the case of the perturbed Kepler problem, they are fixed by injecting the solution in the equation for $(y')^2$ given in~\eqref{eq:yd2}.}. This completes the solution at linear order. Notice that reexpressing $y$ in terms of $\phi$ and subsequently expanding in $\varepsilon$ reproduces the secular terms of the naive perturbative expansion.

\subsubsection{Application to polynomials}

Let us consider the case where the perturbing function is a polynomial $\mathcal{P}\in \mathbb{R}[X]$ of finite degree $n$, written
\begin{equation}\label{eq:P}
\mathcal{P}(y) = \sum_{k=0}^n p_k\, y^k\,.
\end{equation}
As we will see in Sec.~\ref{sec:PN}, it is the most common form of the perturbing function in a post-Newtonian expansion. The first Fourier coefficient of $\mathcal{P}\circ y_0$\footnote{One can also compute it using the Chebyshev reverse formula, see \textit{e.g.} Eq.~(2) of~\cite{MATHAR2006596}.}, which determines $\omega_1$, reads
\begin{equation}\label{eq:a1P}
a_1(\mathcal{P}\circ y_0) = \sum_{k=1}^n k!\, p_k  \sum_{\ell=0}^{\lfloor \frac{k-1}{2} \rfloor}\frac{e^{2\ell+1}}{4^\ell \ell!(\ell+1)!(k-2\ell-1)!}\,.
\end{equation}
Therefore, for a polynomial perturbation, $\omega_1$ is obtained algebraically from the coefficients $p_k$ using Eq.~\eqref{eq:a1P} without evaluating any integrals. Finally, the first order solution~\eqref{eq:y1} becomes a finite sum since $a_{k>n}(g)=0$.

\subsubsection{Application to  logarithms}

Consider a perturbing function that is the product of a logarithm with a polynomial
\begin{equation}\label{eq:pertfP}
f(y) = \ln(y) \mathcal{P}(y)\,.
\end{equation}
The first Fourier coefficient determining $\omega_1$ reads
\begin{equation}\label{eq:a1lnP}
a_1(f\circ y_0) = 2 \sum_{k=0}^n p_k \int_0^{\pi} \frac{\dd \theta}{\pi} \cos(\theta) \left[y_0(\theta)\right]^k\ln y_0(\theta) \,.
\end{equation}
This coefficient is computed using the kernel integrals
\begin{equation}\label{eq:intJk0bis}
\! \! \int_0^{\pi} \!\frac{\dd \theta}{\pi} \cos(k \theta) \ln y_0(\theta) = \!\left\{ 
\begin{array}{ll} 
\ln\left(\frac{1+\sqrt{1-e^2}}{2}\right)&\text{if }k=0\\
-(-\beta)^k/k&\text{if }k\geq 1
\end{array}\right.
\end{equation}
where $\beta=\tfrac{1-\sqrt{1-e^2}}{e}$. The proof is provided in Appendix~A of~\cite{Henry:2026bqh}, where this integral is given by $J_{k,0}(-e)$.

\subsubsection{Application to negative powers}

Suppose $n\geq 1$ and consider the perturbing function
\begin{equation}\label{eq:finv}
f(y) = \frac{1}{y^n}\,.
\end{equation}
The solution can again be constructed from Eq.~\eqref{eq:y1} once the Fourier coefficients of the perturbing function are known. They read
\begin{equation}\label{eq:intakinv}
a_k\left( f\circ y_0\right) = \!\left\{ 
\begin{array}{ll} 
I_{0,n}&\quad\text{if }k=0\\
2(-)^kI_{k,n}&\quad\text{if }k\geq 1
\end{array}\right.\,,
\end{equation}
where the expression of $I_{k,n}$ has been derived in Appendix~A of~\cite{Henry:2026bqh}. For $n\geq 1$, it is given by
\begin{align}
I_{k,n}= \frac{(n+k-1)!}{(n-1)!}\beta^k \sum_{\ell=0}^{n-1}&\frac{1}{2^\ell \ell! (k+\ell)!}\frac{(n+\ell-1)!}{ (n-\ell-1)!} \nn \\
&\times \frac{(1-\sqrt{1-e^2})^{\ell}}{(1-e^2)^{(n+\ell)/2}}\,.
\end{align}

\subsection{Algorithm for higher orders}\label{subsec:algo}

Let us consider the following general equation
\begin{equation}\label{eq:genBinet}
y''+y = 1 + \sum_{k=1}^n \varepsilon^k f_k(y) + \calO(\varepsilon^{n+1}) \,.
\end{equation}
To solve it, we write the generalization of Eqs.~\eqref{eq:yomegalin} as
\begin{subequations}\label{eq:yomegan}
\begin{align}
y(\theta,\varepsilon) &= \sum_{k=0}^n \varepsilon^k y_k(\theta) + \calO(\varepsilon^{n+1})\,,\\
\omega(\varepsilon) &= \sum_{k=0}^n \varepsilon^k \omega_k + \calO(\varepsilon^{n+1})\,.
\end{align}
\end{subequations}
Assuming that we have solved this differential equation at order $n-1$, we know $\omega_k$ and $y_k(\theta)$ for $k\in\llbracket 1,n-1\rrbracket$. Then, injecting~\eqref{eq:yomegan} in~\eqref{eq:genBinet} (recalling that $\dd/\dd\phi = \omega \, \dd/\dd \theta$), we select the $\calO(\varepsilon^n)$ term that takes the form
\begin{equation}
\frac{\dd^2 y_n}{\dd\theta^2} + y_n = \alpha_0^{(n)} + \Bigl[ C_n \omega_n + \alpha_1^{(n)} \Bigr] \cos \theta + \sum_{k\geq 2} \alpha_k^{(n)} \cos(k\theta)\,,
\end{equation}
where $C_n$ and $\alpha_k^{(n)}$ are constants depending on lower-order $\alpha_k^{(m<n)}$ and $(\omega_1,\ldots,\omega_{n-1})$ because of the recursive treatment of the problem. Requiring the coefficient of $\cos\theta$ to vanish fixes $\omega_n=-\alpha_1^{(n)}/C_n$ and guarantees the absence of secular terms. Importantly, $\omega_n$ is therefore determined without solving for $y_n$: the frequency at order $n$ requires the solution $y(\theta)$ only to order $n-1$. Once $\omega_n$ is fixed, $y_n$ can be obtained as in Eq.~\eqref{eq:y1}.

\subsection{Multiple perturbation parameters}\label{subsec:twoscale}

Consider two small parameters $\varepsilon$ and $\varepsilon'$. We want to solve the following differential equation
\begin{equation}\label{eq:PBE2scale}
y''+y=1 + \varepsilon \,f(y) + \varepsilon' \Bigl( g(y) + \varepsilon\, h(y)\Bigr)\,,
\end{equation}
at linear order in $\varepsilon'$ and $\varepsilon$. To do so, we now expand $y$ and $\omega$ in $\varepsilon$ and $\varepsilon'$ as
\begin{subequations}\label{eq:yomegan2}
\begin{align}
y(\theta,\varepsilon,\varepsilon') &= y_{00} + \varepsilon y_{01} + \varepsilon'y_{10} +\varepsilon \varepsilon' y_{11} \,,\\
\omega(\varepsilon,\varepsilon') &= 1 + \varepsilon \omega_{01} + \varepsilon'\omega_{10} +\varepsilon \varepsilon' \omega_{11} \,.
\end{align}
\end{subequations}
By doing so, we get a system of four equations, the ones for $y_{01}$ and $y_{10}$ are of the form~\eqref{eq:d2y1lin}, and are solved in the same way, since they are both leading order in the small parameter. The differential equation for $y_{11}$ is given by
\begin{align}
\frac{\dd^2 y_{11}}{\dd \theta^2}+ y_{11} =&\, h(y_{00}) + y_{01}g'(y_{00}) + y_{10}f'(y_{00}) \\
& -2 \omega_{01}g(y_{00})-2 \omega_{10}f(y_{00})+ 2 y_{10}\omega_{01}\nn\\
& + 2 y_{01}\omega_{10} + 2(y_{00}-1)(\omega_{11}-3\omega_{01}\omega_{10}).\nn
\end{align}
Once $y_{01}$ and $y_{10}$ are solved, we inject their value, together with $\omega_{01}$ and $\omega_{10}$, in the right-hand side. We then find $\omega_{11}$ by canceling the $\cos\theta$ terms and then solve for the differential equation. This allows to fix entirely $\omega$ at a given precision level. This method can also be generalized for higher orders by adapting the algorithm of Sec.~\ref{subsec:algo} as long as the scale separation remains verified.

\section{Post-Newtonian systems}\label{sec:PN}

\subsection{Deriving the perturbed Binet equation}

We consider a binary system of point masses $m_1$ and $m_2$ with $m_1\geq m_2$. In the center-of-mass (CoM) frame, we define the symmetric mass ratio $\nu= m_1 m_2/\tmass^2$ and the normalized mass difference $\delta= (m_1 - m_2)/\tmass$, where the total mass $\tmass = m_1+m_2$. The vector separation from body 2 to body 1 is $\bm{x} = \bm{n} r$, the relative velocity $\bm{v} = \bm{v}_1 - \bm{v}_2$ and the phase angle $\phi$. In the absence of spins, or when the spins are aligned with the orbital angular momentum, the binary's motion remains within a fixed plane. In this case, the conservative motion can be parametrized from the conserved energy $E$ and the norm of the angular momentum $\vert \bm{J}\vert$. We define the reduced quantities
\begin{equation}\label{eq:tildEhdef}
\Et \equiv \frac{E - \tmass c^2}{\tmass \nu}\,, \qquad
\widetilde{L} \equiv \frac{\vert \bm{J}\vert}{\tmass\nu}\,, \qquad h \equiv \frac{\widetilde{L}}{G \tmass}\,.
\end{equation}
When considering a generic perturbation to the Newtonian two-body problem, these on-shell conserved quantities can be written as
\begin{subequations}\label{eq:ELtildeLO}
\begin{align}
\Et &= \frac{\rd^2}{2}+\frac{r^2\fid^2}{2} -\frac{G \tmass}{r} +\varepsilon\, F_E\left(r,\rd,\phi,\fid\right)\,,\\
\widetilde{L} &= r^2\fid + \varepsilon\, F_L\left(r,\rd,\phi,\fid\right)\,.
\end{align}
\end{subequations}
From now on, we suppose that the perturbing functions are independent of $\phi$, i.e. $F_{E,L} =F_{E,L}(r,\rd,\fid)$. This is not always the case, as we will see in Sec.~\ref{subsec:4PN}. The general case is left for future work. To solve this system of equations, we iteratively reduce $\rd$ and $\fid$ in the right-hand sides. The independence in $\phi$ of the perturbing functions implies that the expressions for $\rd^2$ and $\fid$ depend only on~$r$. Defining the dimensionless variable $y = G \tmass h^2/r$, the system~\eqref{eq:ELtildeLO} becomes
\begin{subequations}\label{eq:syst}
\begin{align}
\bigl(y'\bigr)^2 &= 2 \Et h^2 + 2y-y^2+\varepsilon\, F_r(y)\,,\label{eq:yd2}\\
\fid &= \frac{y^2}{G\tmass h^3}+\varepsilon\, F_\phi(y)\,.\label{eq:fidy}
\end{align}
\end{subequations}
Differentiating Eq.~\eqref{eq:yd2} with respect to $\phi$ and factoring out $y'$ yields the perturbed Binet equation (PBE)
\begin{equation}
y''+y = 1 + \varepsilon \, f(y)\,,
\end{equation}
where $f(y)= \tfrac{1}{2} \dd F_r/\dd y$. The PBE takes the desired form solved using the Poincaré–Lindstedt method described in Sec.~\ref{sec:method}. As we will see later, the perturbation parameter $\varepsilon$ depends on the effect we consider. We recall the expression of the Keplerian eccentricity, $e = \sqrt{1+2\Et h^2}$.

\subsection{Periastron advance}

For a perturbed Keplerian orbit, the periastron advance $K$ is defined as the fractional angle between two consecutive passages at periastron. The angle $\theta$, tied to the radial motion, reaches the periastron on a period $2\pi$, thus the periastron advance is given by 

\begin{equation}
K \equiv \frac{\phi(\theta=2\pi)-\phi(\theta=0)}{2\pi}\,.
\end{equation}
Since $\theta = \omega\, \phi$, the periastron advance is related to the frequency introduced in Sec.~\ref{sec:method} through
\begin{equation}\label{eq:Kgen}
K = \frac{1}{\omega} \,.
\end{equation}
This expression is exact. Therefore, the treatment of secular terms in the PBE uniquely determines the periastron advance. At leading order in the perturbation, the periastron advance is then given by
\begin{equation}\label{eq:KLO}
K = 1+\varepsilon \frac{a_1(g)}{2e} + \calO(\varepsilon^2)\,.
\end{equation}
This means that, at linear order in the perturbation, it is sufficient to compute the first Fourier coefficient of the perturbing function $g=f\circ y_0$ to obtain the periastron advance. This can be extended to higher orders using the algorithm of Sec.~\ref{subsec:algo}. 
As emphasized in Sec.~\ref{sec:method}, $\omega_k$, and hence $K$ at that order, is determined before solving for the radial motion $y_k$.

\subsection{Point masses}

Let us first model a binary system of compact objects as two point masses in General Relativity (GR) within a post-Newtonian (PN) expansion, see the Living Review~\cite{Blanchet:2013haa}. The relativistic corrections can then be treated as perturbations of the Kepler problem. The quasi-Keplerian parametrization (QKP), which generalizes the Keplerian parametrization to the PN dynamics, was introduced in Refs.~\cite{DamourDeruelle1,DamourDeruelle2} and subsequently extended to higher PN orders, reaching the 4PN order in~\cite{Cho:2021oai}. In the following, we show that the Poincaré–Lindstedt method reproduces known results of the periastron advance.

\subsubsection{Periastron advance at 3PN}

As described above, we start from the 3PN conserved quantities given in, e.g.~\cite{deAndrade:2000gf}, and derive the corresponding PBE. For illustration, at 1PN, it reads
\begin{equation}\label{eq:PBE1PN}
y'' + y = 1 + \frac{1}{(h c)^2}\left[\Et h^2(4-3\nu) + (6-3\nu) y + \frac{3\nu}{2} y^2\right].
\end{equation}
Setting $\varepsilon_\text{PN} = (hc)^{-2}$, we recognize that the perturbation is a polynomial of $y$. Applying Eq.~\eqref{eq:a1P}, one immediately finds
$\omega_1 = -3$.\footnote{This procedure is also used in standard GR lectures to derive the relativistic periastron advance at 1PN order
.} Notice that at this order, the angle $\theta$ coincides with the true anomaly $v$ in the QKP. This is no longer the case at higher PN orders.

To determine the next frequency correction $\omega_2$, one solves for $y_1$ and applies the recursive procedure of Sec.~\ref{subsec:algo}. At higher PN orders, the degree of the polynomial perturbing the Binet equation increases. At the 3PN order, we find
%
\begin{subequations}\label{eq:omega23}
\begin{align}
\omega_2 =& -\frac{27}{2}+6\nu -e^2\left[\frac{15}{4}-\frac{3}{2}\nu\right]\,,\\
\omega_3 =& -\frac{1647}{16} + \left(\frac{2695}{16}-\frac{123}{32}\pi^2\right)\nu -\frac{21}{8}\nu^2 \nn\\
& + e^2\left[-\frac{435}{8} + \left(\frac{785}{8}-\frac{123}{128}\pi^2\right)\nu -\frac{39}{4}\nu^2\right] \nn \\
& +e^4\left[-\frac{15}{16} + \frac{15}{16}\nu -\frac{3}{4}\nu^2\right]\,.
\end{align}
\end{subequations}
%
Finally, the periastron advance at 3PN is obtained by expanding~\eqref{eq:Kgen} at cubic order
\begin{equation}
K^\text{3PN} = 1 - \frac{\omega_1}{(hc)^2}+ \frac{\omega_1^2-\omega_2}{(hc)^4} - \frac{\omega_1^3-2\omega_1\omega_2+\omega_3}{(hc)^6}
\,.
\end{equation}
Substituting~\eqref{eq:omega23} in this expression, we recover the known results first obtained in Eq.~(20k) of Ref.~\cite{Memmesheimer:2004cv} both in the Arnowitt-Deser-Misner (ADM) and the modified harmonics gauges. See discussion below.

\subsubsection{Comments on the gauges}

The periastron advance, expressed in terms of the conserved energy and angular momentum, is gauge invariant within the class of gauges commonly used in PN and PM calculations. Hence, the present method must yield the same result in these different gauges.

The conserved quantities at the 3PN order have been derived in various gauges. When performing the computation in the usual harmonics coordinates, called standard harmonics (SH), one finds that the perturbing functions $F_E$ and $F_L$ in~\eqref{eq:ELtildeLO}  contain $\ln r$ terms at the 3PN order. However, the usual derivation of the QKP is based on perturbing functions that are polynomial in $1/r$ and involves kernel integrals of the form
\begin{equation}
\int_s^{s_+} \dd x \frac{x^{p-2}}{\sqrt{(s_+-x)(x-s_-)}}\,,
\end{equation}
where $s_+$ and $s_-$ are constants, $s\geq s_-$. An explicit result for all $p\in \mathbb{N}$ is provided in Appendix~A of~\cite{Henry:2025uta}. However, the presence of logarithms in the conserved quantities requires evaluating integrals of the form
\begin{equation}\label{eq:Ipln}
\int_s^{s_+} \dd x \frac{x^{p-2}\ln x}{\sqrt{(s_+-x)(x-s_-)}}\,.
\end{equation}
For $s\neq s_-$, these integrals are considerably more complicated to evaluate. This difficulty is usually avoided by performing a coordinate transformation that removes the logarithmic terms, leading to the so-called modified harmonic (MH) coordinates.

An advantage of the present method is that it only requires the first Fourier coefficient of the perturbing function in the PBE. In the presence of logarithms, one encounters expressions of the form~\eqref{eq:pertfP}, whose first Fourier coefficient can be computed using Eq.~\eqref{eq:a1lnP}. The calculation can therefore be performed directly in SH coordinates, without 
transforming to MH coordinates. We have performed the calculation in SH, MH, and ADM gauges and recover the same result~\eqref{eq:omega23} in all three cases.

Another way of seeing the gauge invariance of $K$ is that the first Fourier coefficient of the difference of the PBE in different gauges $\Delta f$ \textit{must always} vanish. Let us define $\Delta f^\text{harm} = f^\text{MH} (y) - f^\text{SH} (y)$ the difference of the perturbing function between the SH and MH coordinates. It reads
\begin{align}\label{eq:deltaf}
\Delta f^\text{harm} =& \frac{22 y^2 \nu}{3 (hc)^6}\biggl[ 7(1-e^2) -15y + 8y^2 \\
& +\big(12(1-e^2) - 28y + 15 y^2\big) \ln\left(\frac{y \, r'_0}{G\tmass h^2} \right)\biggr]\nn \,, 
\end{align}
where $r'_0$ is a regularization cut-off, see e.g.~\cite{Bernard:2017ktp}. One can indeed check that 
\begin{equation}
a_1(\Delta f^\text{harm}\circ y_0) = 0\,,
\end{equation}
using~\eqref{eq:a1P} and~\eqref{eq:a1lnP}.

\subsubsection{The 4PN contributions}\label{subsec:4PN}

When extending the calculation to the 4PN order, a new difficulty arises due to the conservative tail contribution. This contribution corresponds to a gravitational wave backscattered on the Schwarzschild background and reabsorbed by the binary, see e.g.~\cite{Bernard:2016wrg} for more details. The Hamiltonian can be decomposed as $H = H_0 + H^\text{tail}$, where $H_0$ denotes the ordinary local-in-time 4PN Hamiltonian, while the tail contribution reads
\begin{align}
H^\text{tail} = -\frac{G^2\tmass}{5 c^8}& I_{ij}^{(3)}(T_R) \int_0^\infty  \dd \tau \ln\left( \frac{c\tau}{2\,r(T_R)}\right) \nn \\
&\times \Bigl[ I_{ij}^{(4)}(T_R-\tau) - I_{ij}^{(4)}(T_R+\tau) \Bigr]\,,
\end{align}
where $T_R$ is the retarded time and $I_{ij}^{(n)}$ is the $n^\text{th}$ time derivative of the source mass quadrupole of the binary system. The tail Hamiltonian can in turn be decomposed into a local-in-time contribution $H^\text{tail}_\text{loc}$ and a genuinely nonlocal-in-time contribution $H^\text{tail}_\text{non-loc}$. 

Let us first consider the non-tail contribution $H_0$, which can be treated as before, since it is a polynomial of $y$. We start from the 4PN expressions of $\rd^2$ and $\fid$ of~\cite{Trestini:2025yyc} excluding the tail term. We find that the non-tail contributions at 4PN read
\begin{align}
\omega_4^\text{non-tail} =& -\frac{15039}{16}+\left(\frac{532757}{288} -\frac{121561}{3072}\pi^2 \right)\nu\nn\\
& - \left( \frac{17201}{48} -\frac{5125}{512}\pi^2 \right)\nu^2 -\frac{9}{4}\nu^3 \nn\\
& + e^2 \left[-\frac{2925}{4} + \left(\frac{184151}{96} -\frac{99005}{2048}\pi^2 \right)\nu \right. \nn\\
& \left.\qquad\quad  - \left( \frac{6143}{8} -\frac{4305}{256}\pi^2 \right)\nu^2 +\frac{183}{16}\nu^3 \right] \nn\\
& +e^4 \left[-\frac{3465}{64} + \left(\frac{18703}{96} -\frac{35569}{8192}\pi^2 \right)\nu \right. \nn\\
& \left.\qquad\quad  - \left( \frac{473}{4} -\frac{615}{512}\pi^2 \right)\nu^2 +\frac{81}{8}\nu^3 \right] \nn\\
& - e^6\left( \frac{15}{32} - \frac{3}{8}\nu \right)\nu^2\,.
\end{align}
Substituting this expression into the expansion of Eq.~\eqref{eq:Kgen}, we recover the non-tail 4PN contribution to the periastron advance given in Eq.~(5.8) of~\cite{Bernard:2016wrg} and Eq.~(D6a) of~\cite{Trestini:2025yyc}.

Now, we turn to the local-in-time part of the tail contribution. Since Ref.~\cite{Trestini:2025yyc} computed this contribution separately, we start from the same Hamiltonian given in Eq.~(4.4a) of that reference
%
\begin{align}
H^\text{tail}_\text{loc} &=  \frac{2G^2\tmass}{5 c^8} I_{ij}^{(3)}I_{ij}^{(3)}\ln\left( \frac{r}{\eta}\right)\,,
\end{align}
%
where $\eta$ is a cut-off constant. To derive the associated conserved quantities, we first express $I_{ij}^{(3)}$ in terms of canonical variables, see e.g.  Eq.~(5.15) of~\cite{Damour:2014jta}, and then apply the usual canonical formalism. The on-shell conserved quantities are given by
\begin{subequations}
\begin{align}
\Et^\text{tail}_\text{loc} &= \frac{\rd^2}{2} + \frac{r^2\fid^2}{2} - \frac{G\tmass}{r} \nn \\
& \qquad \quad + \varepsilon\frac{16}{15} \nu y^4\Bigl[\rd^2+12 r^2\fid^2 \Bigr]\ln\left(\frac{y}{y_0'} \right)\,,\\
\Lt^\text{tail}_\text{loc} &= r^2\fid\left( 1 + \varepsilon \frac{128\nu}{5} y^4 \ln\left(\frac{y}{y_0'} \right) \right)\,,
\end{align}
\end{subequations}
where $y_0'=4 \text{e}^{\gamma_E}(1-e^2)^{3/2}/(hc)$ and $\varepsilon = 1/(hc)^8$. Then, we replace perturbatively $\rd$ and $\fid$ to obtain, as before, the associated PBE 
\begin{align}
y''+ y = 1 & + \varepsilon \frac{16\nu}{15} y^3 \biggl[ 23(e^2-1)+46y-11y^2 \\
& + 2 \bigl(46(e^2-1)+115y-33 y^2\bigr)\ln\left(\frac{y}{y_0'}\right)\biggr] \,.\nn
\end{align}
This perturbing function has the same structure as~\eqref{eq:deltaf}, namely a polynomial together with logarithmic terms. Its first Fourier coefficient is directly obtained using~\eqref{eq:a1P} and~\eqref{eq:a1lnP}. The resulting periastron advance reads 
\begin{align}
K_\text{loc}^\text{tail} = 1 & + \frac{\nu}{90 e^2 (hc)^8} \Biggl[ 1152\Bigl(1-\sqrt{1-e^2} \Bigr) \nn\\
& + 8 e^2 \Bigl(3637-2917\sqrt{1-e^2}\Bigr)  \nn\\
& + 4 e^4\Bigl(5982-2803\sqrt{1-e^2}\Bigr)+1209e^6 \nn\\
& +12e^2\bigl(1256+1608e^2+111e^4\bigr)\nn\\
& \quad\times\left( \ln (hc) -\gamma_E +\ln \frac{1+\sqrt{1-e^2}}{8(1-e^2)^{3/2}}\right) \Biggr]\,.
\end{align}
We recover exactly the result~(D6b) of~\cite{Trestini:2025yyc}.

Regarding the purely non-local-in-time terms, the perturbing functions $F_{E,L}$ of~\eqref{eq:ELtildeLO} depend on $\phi$. We cannot directly apply the method of Sec.~\ref{sec:method} since the perturbing function of the PBE depends as well on time, which is linked to the mean anomaly through $\ell = n(t-t_0)$, where $n$ is the mean motion. Symbolically,
\begin{equation}
y'' + y = 1 + \varepsilon\, f\bigl(y,\ell(\phi)\bigr)\,.
\end{equation}
When dealing with such equation, the perturbing function is not necessarily even anymore. Thus, the coefficients $b_n$ of the Fourier series do not vanish. In this case, the complex formulation of Fourier series may be more adapted. The main difficulty when computing the tail integrals is the presence of functions of $T_R\pm\tau$. To evaluate them, one must invert $\phi(\ell)$ from the Kepler equation
\begin{equation}
\ell = 2\arctan\left[\sqrt{\frac{1-e}{1+e}}\tan\frac{\phi}{2} \right] -e\sqrt{1-e^2}\frac{\sin\phi}{1+e\cos\phi}\,.
\end{equation}
This can be done using Fourier series as well. When computing the first Fourier coefficient of the perturbing function, one would end up on an infinite series of Hansen coefficients that are, a priori, not resummable in a closed form. This computation is left for future work.

\subsection{Tidal effects}

Tidal interactions in a binary system can also be treated as perturbations of the Kepler problem. The deformation of extended bodies in an external gravitational field can be described through a multipole expansion~\cite{Bini:2012gu}. Here we consider the adiabatic-tides model, both in Newtonian gravity and in GR, in which the induced multipole moments respond instantaneously to the tidal field generated by the companion. Each body $A$ is characterized by tidal polarizabilities $\mu^{(\ell)}_A$, which parametrize its deformability and are related to its tidal Love numbers; see, e.g., Eq.~(2.5) of~\cite{Bini:2012gu}. We define the following combinations
\begin{equation}\label{eq:mutilde}
\mu_+^{(\ell)} = \frac{m_2}{2m_1}\mu_1^{(\ell)} + \frac{m_1}{2m_2}\mu_2^{(\ell)}, \quad \widetilde{\mu}_+^{(\ell)} = G \mu_+^{(\ell)} \left( \frac{c^2}{G\tmass}\right)^{2\ell+1}.
\end{equation}
For black holes, the value of the individual tidal polarizabilities vanish identically~\cite{Damour:2009vw,LeTiec:2020bos}.

\subsubsection{Adiabatic tides in Newtonian gravity}\label{subsubsec:tidesNewt}

In Appendix~A of~\cite{HFB19}, we derived the Lagrangian in Newtonian gravity modeling adiabatic tides. Here, we restrict ourselves to terms linear in tidal polarizabilities. In the center-of-mass (CoM) frame, the contribution of the $\ell^\text{th}$ mass-type tidal multipole to the Lagrangian reads
\begin{equation}
\frac{\Lag_{\mu^{(\ell)}}}{\tmass \nu} = \frac{v^2}{2} +\frac{G\tmass}{r} +  (2\ell-1)!!\frac{G^2\tmass \mu_+^{(\ell)}}{r^{2\ell+2}}\,.
\end{equation}
The corresponding conserved quantities are
\begin{subequations}
\begin{align}
\Et_{\mu^{(\ell)}} &= \frac{\rd^2}{2} + \frac{r^2\fid^2}{2} - \frac{G\tmass}{r} - (2\ell-1)!! \frac{G^2\tmass \mu_+^{(\ell)}}{r^{2\ell+2}}\,,\\
\Lt_{\mu^{(\ell)}} &= r^2\fid\,.
\end{align}
\end{subequations}
The associated PBE is given by
\begin{equation}\label{eq:PBEtides}
y'' + y = 1 + \varepsilon_{\mu^{(\ell)}} y^{2\ell+1}\,,
\end{equation}
with 
$\varepsilon_{\mu^{(\ell)}} = 2 (\ell+1)(2\ell-1)!! \widetilde{\mu}^{(\ell)}_+/(hc)^{4\ell+2}$. In Sec.~II of~\cite{Henry:2025uta}, we applied the present method to the leading tidal interaction ($\ell=2$) and showed that $\varepsilon_{\mu^{(2)}}$ is very small for realistic compact binaries. Since the perturbation in~\eqref{eq:PBEtides} is a monomial, its contribution to the periastron advance follows directly from Eq.~\eqref{eq:a1P}
\begin{equation}\label{eq:Kmul}
K_{\mu^{(\ell)}} = 1+\frac{\widetilde{\mu}_+^{(\ell)}}{(hc)^{4\ell+2}}\sum_{k=0}^\ell\frac{(\ell+1)(2\ell)!(2\ell+1)!!}{4^k k! (k+1)!(2\ell-2k)!}e^{2k}.
\end{equation}
In 1939, Sterne derived the contributions up to $\ell=4$, displayed in Eqs.~(14)-(16) of~\cite{Sterne1939}. For comparison, we recall that the normalized angular momentum is linked to the semi-major axis $a$ through $h^2 = (1-e^2)a/(G\tmass)$. These results, together with the results obtained with a QKP method in~\cite{Henry:2025uta,Henry:2026bqh} for $\ell =2$ and~3 are in agreement. The general expression for arbitrary $\ell$ is new.

\subsubsection{Adiabatic tides in GR: Current multipoles at LO}

In GR, the effective action describing adiabatic tidal interactions has been derived in Refs.~\cite{Damour:1990pi,Damour:1991yw,Damour:1992qi}. An explicit expression, written in a form analogous to the Newtonian action, can be found, e.g., in Eq.~(2.3) of~\cite{Bini:2012gu}. Here we restrict ourselves to terms linear in the tidal polarizabilities. In addition to the mass-type tidal moments already present in Newtonian gravity, the relativistic action contains current-type tidal moments, parametrized by the rotational tidal polarizabilities $\sigma^{(\ell)}_A$. We consider their leading-order contribution for arbitrary $\ell$. The corresponding Lagrangian, derived in Appendix~\ref{app:Lsigma}, reads
\begin{equation}\label{eq:Lsigma}
\frac{\Lag_{\sigma^{(\ell)}}}{\tmass \nu} = \frac{v^2}{2} +\frac{G\tmass}{r} + D_\ell\frac{G^2\tmass \sigma_+^{(\ell)}}{c^2 r^{2\ell+2}}\bigl(v^2-(nv)^2\bigr)\,,
\end{equation}
where $D_\ell = 8(2\ell-1)!!$. The corresponding conserved energy and angular momentum are
\begin{subequations}
\begin{align}
\Et_{\sigma^{(\ell)}} &= \frac{\rd^2}{2} + \frac{r^2\fid^2}{2} - \frac{G\tmass}{r} + D_\ell \frac{G^2\tmass \sigma_+^{(\ell)}}{c^2}\frac{\fid^2}{r^{2\ell}}\,,\\
\Lt_{\sigma^{(\ell)}} &= r^2\fid\left( 1 + 2 D_\ell \frac{G^2\tmass \sigma_+^{(\ell)}}{c^2 r^{2\ell+2}}\right)\,.
\end{align}
\end{subequations}
The fact that the current multipoles contribute to the angular momentum makes the PBE more complex. Indeed, it reads
\begin{align}
y'' + y = 1 + \varepsilon_{\sigma^{(\ell)}} \Bigl[& 2\Et h^2 (\ell+1)y^{2\ell+1} +(2\ell+3) y^{2\ell+2}\nn \\
& -\frac{\ell+2}{2}y^{2\ell+3} \Bigr]\,,
\end{align}
with 
$\varepsilon_{\sigma^{(\ell)}} = 32(2\ell-1)!!\widetilde{\sigma}_+^{(\ell)}/(hc)^{4\ell+4}$, where $\widetilde{\sigma}_+^{(\ell)}$ is obtained by replacing $\mu$ by $\sigma$ in~\eqref{eq:mutilde}.
Once again, the perturbing function is a polynomial and one simply injects~\eqref{eq:a1P} in~\eqref{eq:KLO}, to obtain
\begin{small}
\begin{align}\label{eq:Ksigmal}
K_{\sigma^{(\ell)}} =&\, 1+ \frac{8(2\ell+3)!(2\ell-1)!!\widetilde{\sigma}_+^{(\ell)}}{(hc)^{4\ell+4}}\Biggl[ - \frac{e^{2\ell+2}}{4^{\ell+1}[(\ell+1)!]^2} \nn\\
& + \sum_{k=0}^\ell \frac{e^{2k}}{4^k k!(k+1)!} \biggl( \frac{e^2-1}{(2\ell+3)(2\ell-2k)!} \nn\\
& +\frac{2}{(2\ell-2k+1)!}-\frac{\ell+2}{(2\ell-2k+2)!} \biggr)\Biggr]\,.
\end{align}
\end{small}
For $\ell=2$, Eq.~\eqref{eq:Ksigmal} reproduces the LO result Eq.~(B4a) of Ref.~\cite{Henry:2025uta}. To our knowledge, the general result for arbitrary $\ell$ is new. For $\ell=3$, we also find agreement for the coefficient of $h^{-8}$ with the PN expansion of the post-Minkowskian (PM) result of Ref.~\cite{Kalin:2020lmz}. We discuss such comparison in more detail in the following subsection.

\subsubsection{Mass-type tidal quadrupole to NNLO}\label{subsec:MQNNLO}

From the effective relativistic action mentioned above, one can derive the tidal corrections to the Lagrangian and/or the Hamiltonian in different gauges. In~\cite{HFB19,HFB20b}, we computed the tidal mass quadrupole to NNLO, the current quadrupole to NLO and the mass octupole to LO in harmonic coordinates, ADM and isotropic coordinates. Starting from the associated Lagrangian or Hamiltonian, we can derive the conserved quantities available in Eqs.~(5.5) and (5.6) of~\cite{HFB19}. Then, as before, it is straightforward to deduce the PBE including the PN corrections to the tidal terms. It symbolically takes the form
\begin{align}\label{eq:NNLOmu2}
y'' + y =& \, 1 + \varepsilon_\text{PN}\, f_1(y) + \varepsilon_\text{PN}^2 \,f_2(y) \nn\\
& + \varepsilon_{\mu^{(2)}}\Bigl( g_0(y) +  \varepsilon_\text{PN} \, g_1(y)+  \varepsilon_\text{PN}^2 \, g_2(y)\Bigr)\,,
\end{align}
where $f_k$ are the point-mass PN corrections ($f_1$ is explicited in~\eqref{eq:PBE1PN}) and $g_k$ are the tidal corrections. This is a two-scale expansion since $\varepsilon_{\mu^{(2)}}\ll \varepsilon_\text{PN}$~\cite{Henry:2025uta}, where the general case is treated in Sec.~\ref{subsec:twoscale}. We have applied it to the mass quadrupole to NNLO and we recover exactly the result derived from the QKP method displayed in Eq.~(B4a) of~\cite{Henry:2025uta}.

\subsubsection{Current-type tidal quadrupole to NNLO}

In~\cite{HFB19}, the current-type quadrupole contributions to the Lagrangian have been derived up to NLO, i.e. 6PN and 7PN. This interaction is parametrized by the current-type tidal multipole  $H_L$ (it is the magnetic part of the Weyl tensor, see e.g.~(2.3b) of~\cite{HFB19} for a proper definition). Notably, we derived the tidal invariant $(H_L^A)^2$ to NLO off shell, i.e. without replacing the accelerations by the equations of motion, in order to control the gauge. If one performs this replacement in the Lagrangian, the gauge is changed. On the other hand, in~\cite{Dones:2024odv}, we derived the tidal current quadrupole tensor on shell to NNLO. Since the periastron advance is gauge invariant, the results with the on-shell and off-shell values must be equivalent. Thus, we derived the Lagrangian from the on-shell value of the tidal invariant in a gauge that is not controlled. Next, we obtained the conserved quantities to NNLO in this gauge, and derived the PBE which takes the same form as~\eqref{eq:NNLOmu2}. By applying the exact same method, we find the contributions of the current quadrupole deformation to the periastron advance to NNLO
\begin{widetext}
\begin{align}
K_{\sigma^{(2)}} =& \, 1 + \frac{\sigmatp}{2(hc)^{12}}\Bigl( 1248 + 4320 e^2 + 1980 e^4 + 75e^6\Bigr)  + \frac{1}{(hc)^{14}}\Biggl\{  \frac{\sigmatp}{16} \Bigl[ 64\bigl(5746-189\nu\bigr) + 48\bigl(25453 - 3619\nu\bigr) e^2 \nn\\
& + 30 \bigl(25507-7665\nu\bigr) e^4 + \bigl(92630 - 51660\nu\bigr) e^6  + 75 \bigl(13-14\nu\bigr) e^8\Bigr] +\frac{17\delta\,\sigmatm}{8}\Bigl( 544 + 2688 e^2 + 2100 e^4 + 245 e^6\Bigr) \Biggr\}\nn\\
& + \frac{1}{(hc)^{16}}\Biggl\{  \frac{\sigmatp}{1344} \Bigl[ 8\bigl(91752179- 15408346\nu - 308133\nu^2\bigr) + 14\bigl(177440539 - 64123682\nu + 1300275\nu^2\bigr) e^2  \nn\\
& \qquad \qquad + 42 \bigl(45779551 - 27752207\nu + 2063952\nu^2\bigr) e^4 + 70\bigl(5588033 - 5180422\nu + 853002\nu^2\bigr) e^6  \nn\\
& \qquad \qquad + 35 \bigl(476057 - 658510\nu + 229464\nu^2\bigr) e^8  + 3150 \bigl(17 - 39\nu + 33\nu^2\bigr) e^{10}\Bigr] \nn\\
& \qquad +\frac{\delta\,\sigmatm}{1344}\Bigl[ 64\bigl(1072975- 19992\nu\bigr) + 560\bigl(556255 - 62373\nu\bigr) e^2 + 42 \bigl(7066075- 1689681\nu\bigr) e^4  \nn\\
& \qquad \qquad + 70\bigl(969161 - 410550\nu\bigr) e^6 + 245 \bigl(10889- 7854\nu\bigr) e^8\Bigr] \Biggr\}\,.
\end{align}
\end{widetext}
This expression agrees with Eq.~(B4a) of~\cite{Henry:2025uta} up to NLO, while the NNLO term is new. As an independent check, we compare our results with the PM calculations of Refs.~\cite{Cheung:2020sdj,Kalin:2020lmz}. Using the boundary-to-bound mapping of Ref.~\cite{Kalin:2019rwq}, we reconstruct the periastron advance from the PM scattering angle and expand it in the PN regime. At the available PM order, this provides an independent check of the NNLO $\calO(e^8)$ coefficient for the mass quadrupole and the NNLO $\calO(e^{10})$ coefficient for the current quadrupole. We find perfect agreement in both cases.

\subsection{Spins}

The action modeling spins of compact objects in GR can also be computed from effective field theory. Several works tackled this problem, notably~\cite{Marsat:2014xea,Levi:2015msa,Porto:2016pyg}. In this approach, there are two main types of interactions: spin-orbit and spin-induced multipoles. The bodies are endowed with spin vectors $\bm{S}_A$ of constant norm. We define the combinations
\begin{align}
\bm{S} &= \bm{S}_1 + \bm{S}_2\,,\\
\bm{\Sigma} &= \frac{\tmass}{m_2} \bm{S}_2 - \frac{\tmass}{m_1} \bm{S}_1 \,.
\end{align}
We restrict to configurations in which the spins are aligned with the orbital angular momentum, so that the motion remains within a fixed plane. Thus, we consider their projection perpendicular to the orbital plane $S_l = (\bm{S}.\bm{l})$ and $\Sigma_l = (\bm{\Sigma}. \bm{l})$. Finally, we define the dimensionless quantities
\begin{equation}
s_l = \frac{S_l}{G \tmass^2}\,, \qquad \sigma_l =  \frac{\Sigma_l}{G \tmass^2}\,.
\end{equation}

\subsubsection{Spin-orbit to NNLO}

The conserved quantities including the spin-orbit (SO) interaction have been derived in~\cite{Marsat:2012fn,Bohe:2012mr} in harmonic coordinates up to the 3.5PN order corresponding to the NNLO. Although also conserved, the orbital angular momentum $\bm{L}$ differs from the total angular momentum $\bm{J}$ by $\bm{L}=\bm{J}-\bm{S}/c$. At leading order, the conserved quantities that we start from are given by
\begin{subequations}
\begin{align}
\Et_\text{SO} &= \frac{\rd^2}{2} + \frac{r^2\fid^2}{2} - \frac{G\tmass}{r} +\frac{G(S_l +\delta\,\Sigma_l)}{c^3} \frac{\fid}{r}\,,\\
\Lt_\text{SO} &= r^2\fid\ - \frac{G(3 S_l +\delta\,\Sigma_l)}{r c^3} + \frac{(S_l +\delta\,\Sigma_l)(\rd^2+r^2\fid^2)}{2\tmass c^3}\,.
\end{align}
\end{subequations}
As usual, we derive the PBE, which reads at LO
\begin{equation}
y'' + y = 1 - \frac{1}{(hc)^3}\biggl[ 2\Et h^2(s_l-\delta\sigma_l) +8 s_l\, y + 3(s_l+\delta\sigma_l)y^2 \biggr]\,.
\end{equation}
We compute the linear-in-spin contributions to NNLO (3.5PN) using the two-parameter expansion of Sec.~\ref{subsec:twoscale}. The periastron advance is given by
\begin{align}
K_\text{SO} =& \, 1 - \frac{7 s_l + 3\delta\sigma_l}{(hc)^3} \nn \\
& - \frac{1}{2(hc)^5} \biggl[ \Bigl(237-25\nu+e^2(36-14\nu)\Bigr)s_l\nn \\
& \qquad +3 \delta\Bigl(31-4\nu + 2e^2(2-\nu)\Bigr)\sigma_l \biggr]\nn\\
& - \frac{3}{8(hc)^7}\biggl[ \Bigl(4615-1777\nu+5\nu^2  + e^2( 1630  \nn\\
& \qquad - 1260\nu + 76\nu^2) + e^4( 25- 38\nu + 14\nu^2) \Bigr)s_l\nn \\
& \qquad + \delta \Bigl(1755-788\nu + 3\nu^2 + e^2( 550- 520\nu \nn \\
& \qquad + 36\nu^2) + e^4( 5- 12\nu + 6\nu^2) \Bigr)\sigma_l \biggr]\,,
\end{align}
which is in agreement with the literature~\cite{Tessmer:2012xr,LeTiec:2013uey,Henry:2023tka}.

\subsubsection{Mass-type spin-induced multipoles}

The effective Lagrangian modeling spin-induced nonminimal couplings is given in Eq.~(4.16) of~\cite{Levi:2015msa}. In this section, we focus on the even-in-spin interactions, which act as a mass-type multipole. At LO, after reestablishing the $G$ and $c$ powers, it reads
\begin{equation}
L_{S^{2\ell}} = \frac{(-)^\ell}{(2\ell)!} \frac{C_{ES^{2\ell}}^1}{c^{4\ell} m_1^{2\ell-1}}G_{2L}^1 S^1_{2L} +1\leftrightarrow 2\,,
\end{equation}
where $G_L$ is the mass-type tidal multipole (it is the electric part of the Weyl tensor, once again see e.g.~(2.5a) of~\cite{HFB19}) and $C_{ES^{2\ell}}^A$ are the Wilson coefficients associated with the $\ell^\text{th}$ mass-type spin-induced multipole deformation of body $A$. For black holes, these coefficients are unity, while for other compact objects they depend on the equation of state. Substituting the mass-type tidal tensor $G_{2L}$ with~(2.6a) of~\cite{HFB19}, we find the Lagrangian
\begin{align}\label{eq:LS2l}
\frac{\Lag_{S^{2\ell}}}{\tmass \nu} =&\, \frac{v^2}{2} +\frac{G\tmass}{r} + \frac{(-)^\ell G}{r^{2\ell+1} c^{4\ell} \tmass^{4\ell-1}\nu^{2\ell}}\nn\\& \times \left[ m_2^{2\ell} |S_1|^\ell C_{ES^{2\ell}}^1 P_{2\ell} \left( \frac{n.S_1}{|S_1|} \right) + 1\leftrightarrow 2\right]\,,
\end{align}
where $P_{N}$ is the usual Legendre polynomial. The conserved quantities, after imposing that the spins are perpendicular to the orbital plane, $(n.S_A)=0$, read
\begin{subequations}
\begin{align}
\Et_{S^{2\ell}} &= \frac{\rd^2}{2} + \frac{r^2\fid^2}{2} - \frac{G\tmass}{r} - \frac{\kappa^{(2\ell)}}{r^{2\ell+1}}\,,\\
\Lt_{S^{2\ell}} &= r^2\fid\,,
\end{align}
\end{subequations}
where $\kappa^{(2\ell)}$ is the last term of~\eqref{eq:LS2l} after substituting $P_{2\ell}(0) = (-)^\ell\tfrac{(2\ell-1)!!}{(2\ell)!!}$. The corresponding PBE is
\begin{equation}
y'' + y = 1 + \varepsilon_{S^{2\ell}} \, y^{2\ell}\,,
\end{equation}
where
\begin{equation}
\varepsilon_{S^{2\ell}} = \frac{1}{(h c)^{4\ell}}\frac{(2\ell+1)!!}{(2\ell)!!} \frac{m_2^{2\ell} |S_1|^\ell C_{ES^{2\ell}}^1 
}{(G\tmass^3\nu)^{2\ell}}  + 1\leftrightarrow 2 \,.
\end{equation}
Using~\eqref{eq:a1P} and~\eqref{eq:KLO}, the periastron advance is given by
\begin{equation}
K_{S^{2\ell}} = 1+\frac{\varepsilon_{S^{2\ell}}}{2}\sum_{k=0}^{\ell-1} \frac{(2\ell)!}{4^k k! (k+1)!(2\ell-2k-1)!}e^{2k}.
\end{equation}
This formula yields the correct result for the spin-induced quadrupole ($\ell=1$) given in Eq.~(A5b) of Ref.~\cite{Henry:2023tka} at NLO. To our knowledge, this formula for higher multipoles is new.

\subsection{Electromagnetic interaction}

So far, we have only considered gravitational effects perturbing the Kepler problem. Now, we allow the binary to be electromagnetically interacting. We first consider the electromagnetic dipole interaction, and then turn to the charged-particles case.

\subsubsection{Electric and magnetic dipoles}

In~\cite{Henry:2023guc}, we considered two massive bodies endowed with electromagnetic dipoles interacting through the Einstein-Maxwell action in order to model a gravitationally bound system in magnetic interaction. We notably derived the Lagrangian and associated Noetherian quantities, including the electric dipoles $\bm{q}_A$ to NNLO in the PN expansion, and the magnetic dipoles $\bm{J}_A$ to NLO. Their explicit expressions are available in Appendix~A of~\cite{Henry:2023guc}. Similarly to the spins, we assume in this section that the dipoles are perpendicular to the orbital plane so that it remains constant. As for the spins, we define the dimensionless electric and magnetic dipoles
\begin{equation}
\tilde{q}_A = \frac{(\bm{q}_A.\bm{l})}{G \tmass^2}\,, \qquad \tilde{\mu}_A =  \frac{(\bm{J}_A.\bm{l})}{G \tmass^2}\,,
\end{equation}
See Sec.~II~B of~\cite{Henry:2023guc} for proper definitions of the dipoles. 

In~\cite{Henry:2023len}, we derived the QKP for the magnetic dipole terms to NLO, displayed in Eqs.~(4.8). This was done in order to model binary neutron star systems or double white dwarfs, that are expected to bear strong magnetic dipoles. With the Poincaré method, we find their contributions at  NLO to the periastron advance
\begin{equation}
K_{J^2} = 1 - \frac{3\alpha \tilde{\mu}_1\tilde{\mu}_2}{\nu (hc)^4} \Biggl[ 1 + \frac{38 + 2\nu +e^2(7-2\nu) }{2(hc)^2} \Biggr],
\end{equation}
where $\alpha$ is the fine structure constant. It is in agreement with Eq.~(B10c) of~\cite{Henry:2023len}.

Although not physically relevant, because we do not expect realistic compact objects to bear constant electric dipoles, we derive their contributions to the periastron advance to NNLO. Starting from the Noetherian quantities of~\cite{Henry:2023guc}, which become conserved when assuming the electric dipoles to be aligned with the orbital angular momentum, we employ the multi-scale method of Sec.~\ref{subsec:twoscale} where the small parameters are $\varepsilon_\text{PN}$ and $\alpha$. The periastron advance can be split between $q_1 q_2$ and $q_A^2$ terms, namely $K_{q^2} = 1 + k_{q_1 q_2} + k_{q_1^2} + k_{q_2^2}$. The cross terms read
\begin{align}
k_{q_1 q_2} =&\, - \frac{3\alpha \tilde{q}_1 \tilde{q}_2}{\nu c^2 h^4} \Biggl[ 1 +   \frac{30 + 2\nu +e^2(5-2\nu)}{2(hc)^2}  \\
& + \frac{1}{24(hc)^4}\biggl(4991-409\nu-86\nu^2 + 6 e^2\Bigl(302  \nn \\
& -145\nu - 28\nu^2 \Bigr) + 3 e^4\Bigl(5-24\nu +3\nu^2 \Bigr)\biggr)\Biggr]\,,\nn
\end{align}
and the self terms are purely NNLO contributions
\begin{align}
k_{q_1^2} = \frac{\alpha\,\tilde{q}_1^2}{448\nu c^6 h^8} \biggl[& 8 \bigl(77-38\nu - 7\delta(11+26\nu) \bigr) \\
& +12 e^2\bigl(49+2\nu- 7\delta(7 + 22\nu)\bigr) \nn\\
& + 21e^4\bigl(1+2\nu -\delta(1+6\nu)\bigr)\biggr]\,,\nn
\end{align}
where $k_{q_2^2}$ is obtained by replacing $\{q_1,\delta\}\rightarrow \{q_2,-\delta\}$. To our knowledge, this is a new result at all PN orders.

\subsubsection{Electric charge to NNLO}

In~\cite{Placidi:2025xyi}, the authors derived the conservative motion of a binary system in which both particles have an electric charge to the NNLO. They notably published the periastron advance in the circular limit. In this section, starting from the Lagrangian available in their ancillary file, we derived the conserved quantities to 2PN. We define $\rho = (1-\eta_1 \eta_2)$ where $\eta_A$ are dimensionless electric charges defined in Sec.~II of~\cite{Placidi:2025xyi}. We display here the 1PN result
\begin{subequations}
\begin{align}
\Et_c =&\, \frac{\rd^2}{2} + \frac{r^2\fid^2}{2} - \frac{G\tmass\rho}{r}  + \frac{1}{c^2} \biggl[ \frac{3(1-3\nu)}{8}(\rd^2 +r^2\fid^2)^2 \nn\\
& + \frac{G\tmass}{2r}\Bigl( (3+2\rho\nu)\rd^2 +(3+ \rho\nu)r^2\fid^2\Bigr)\\
& + \frac{G^2\tmass^2}{4r^2}\Bigl( 2-4\eta_1 \eta_2+(1+\delta)\eta_1^2+(1-\delta)\eta_2^2 \biggr] \nn\,,\\
\Lt_c =&\, r^2\fid \left[ \frac{1-3\nu}{2}(\rd^2 +r^2\fid^2)+\frac{G\tmass}{r}\bigl(3+\nu\rho\bigr)\right],
\end{align}
\end{subequations}
Notice that here, the leading order electric charge actually acts as a correction to the mass, thus the intermediate computations are done employing the rescaled quantities
\begin{equation}
\widetilde{m}= \rho \, \tmass ,\quad \tilde{h}=\frac{|\bm{J}|}{G \tmass \tilde{m}\nu},\quad y = \frac{G \widetilde{m} \tilde{h}^2}{r}.
\end{equation}
With these definitions, we recover the structure of the equations of Sec.~\ref{sec:method}, yielding the rescaled eccentricity $\tilde{e}^2=1+2\Et\tilde{h}^2$, and we can directly apply the method of Sec.~\ref{sec:method}. We derived the PBE to 2PN, we display here only the 1PN
\begin{align}\label{eq:PBEcharge}
y'' + y =&\, 1 + \frac{1}{(hc)^2}\Bigl[ \frac{\Et h^2}{\rho} \Bigl(4-3\nu - \eta_1 \eta_2(1-3\nu)\Bigr) \nn \\
& + \frac{y}{2} \Bigl( 6(2-\nu) -12 \eta_1 \eta_2(1-\nu) -\eta_1^2 - \eta_2^2 \nn \\
& +2\eta_1^2\eta_2^2(1-3\nu) - \delta(\eta_1^2-\eta_2^2)\Bigr)+ \frac{3\nu\rho^2}{2} y^2\Bigr].
\end{align}
Then, we deduce the periastron advance to 2PN
\begin{align}\label{eq:Kc}
K_c =&\, 1 + \frac{12-12 \eta_1 \eta_2 -\eta_1^2 - \eta_2^2+2\eta_1^2\eta_2^2 - \delta(\eta_1^2-\eta_2^2)}{4(hc)^2}\nn\\
& + \frac{1}{16(hc)^4}\Bigl[ 210-60\nu + 12 \Et h^2(5-2\nu)\nn\\
& - 4 \eta_1 \eta_2\bigl(6\Et h^2 (1-\nu) +15(7-3\nu)\bigr)\nn\\
& -\eta_1^2\Bigl((1+\delta)\bigl(90+ 4\Et h^2(3-\nu)\bigr)+6\nu(1-\delta)\Bigr) \nn\\
& + \eta_1^2\eta_2^2\Bigl( 270-4\bigl(54+\Et h^2\bigr)\nu \Bigr) \nn\\
& -6\eta_1^2\eta_2^4 \bigl( 6+5\nu-\delta(6-\nu)\bigr) \nn\\
& + 12 \eta_1\eta_2^3\bigl( 10+3\nu -\delta(10-\nu) \bigr)-60 \eta_1^3\eta_2^3(1- 2\nu) \nn\\
& +3 \eta_1^4(1-2\nu+\delta)+\eta_1^4\eta_2^4(3-18\nu) + 1\leftrightarrow 2 \Bigr].
\end{align}
The periastron advance was computed in~\cite{Placidi:2025xyi} to 2PN in the circular limit. We have checked that our result agrees with theirs in that limit. To do so, we have expressed both energy and angular momentum in terms of the orbital frequency and substituted them in~\eqref{eq:Kc}. The result in the general (eccentric) case is in agreement with Eq.~(81) of~\cite{Alonzo-Artiles:2026wbe}.

As a consistency check, the periastron advance can be derived from the PBE obtained from the geodesic equation for a charged test mass in the Reissner–Nordström metric. It has been derived in Ref.~\cite{Das:2016opi} and is provided in Eq.~(52). In our notations it reads
\begin{align}
y'' + y = 1 &+  \frac{1}{(hc)^2} \Biggl[ 3 \rho\,y^2 -\eta_1^2(1-\eta_2^2)y  - \frac{\eta_1 \eta_2\Et h^2}{\rho} \Biggr] \nn \\
& -\frac{2\eta_1^2\rho^2}{(hc)^4}  y^3\,.
\end{align}
It differs from~\eqref{eq:PBEcharge} in the test-mass limit $\nu = 0$ and $\delta = 1$. However, the associated periastron advance at second order yields the test-mass limit of Eq.~\eqref{eq:Kc}.

\subsection{Test particle in de Sitter-Schwarzschild metric}

In~\cite{Alvarez-Perez:2025pxi}, the author derived the Binet equation for a massive test particle in the de Sitter-Schwarzschild metric. Writing their Eq.~(31) in our notations, we have
\begin{equation}
y''+ y = 1 + 3\, \varepsilon\, y^2 - \frac{\varepsilon'}{3} y^{-3}\,,
\end{equation}
where $\varepsilon = (hc)^{-2}$ and the dimensionless parameter $\varepsilon'$ is linked to the cosmological constant via $\varepsilon' = \Lambda (G\tmass c\, h^3)^2$. At linear order in $\varepsilon'$, we can use the formula~\eqref{eq:intakinv} to obtain the periastron advance
\begin{align}
K_\Lambda &= 1 + \frac{\varepsilon'}{2(1-e^2)^{5/2}} + \calO(\varepsilon,\varepsilon'^2)\nn\\
&= 1 + \frac{\Lambda a^3 c^2}{2G\tmass}\sqrt{1-e^2}+\calO(\varepsilon,\Lambda^2)\,,
\end{align}
where $a = - G\tmass/(2\Et) = G\tmass h^2/(1-e^2)$ is the Newtonian semi-major axis. This result is in agreement with the literature, see e.g. Eq.~(38) of Ref.~\cite{Kerr:2003bp}.

\subsection{Perturbation in a Yukawa potential}

To finish, we tackle the case where the binary is submitted to a Yukawa potential interaction
\begin{equation}
V(r) = -\frac{G\tmass}{r} \left( 1+ \alpha \de^{-r/\lambda} \right)\,,
\end{equation}
where $\alpha > 0$ represents the strength of interaction and $\lambda$ a scale parameter. Such a potential yields the PBE
\begin{equation}
y''+ y = 1 + \alpha\left( 1 + \frac{\rho}{y} \right)\de^{-\rho/y} \,,
\end{equation}
where $\rho = G\tmass h^2/\lambda$. In this case, the first Fourier coefficient associated with that perturbation is more complicated. It reads
\begin{equation}
a_1(f\circ y_0) = \frac{2}{\pi}\int_0^\pi \dd \theta\cos\theta \left( 1+\frac{\rho}{1+e \cos\theta}\right)\de^{-\frac{\rho}{1+e \cos\theta}}\,.
\end{equation}
This integral can be evaluated in a closed form using the Weierstrass substitution $\tan\tfrac{\theta}{2} = \sqrt{\tfrac{1+e}{1-e}}\tan \tfrac{u}{2}$ and an integration by part. Substituting its value in Eq.~\eqref{eq:KLO}, we get at leading order in $\alpha$
\begin{equation}
K_\text{Yukawa} = 1 + \alpha \frac{a\sqrt{1-e^2}}{\lambda e}\de^{-a/\lambda} I_1\left( \frac{ae}{\lambda}\right) + \calO\bigl(\alpha^2\bigr)\,,
\end{equation}
where $I_1$ is the modified Bessel function of the first kind, and $a$ the Newtonian semi-major axis. This expression is exact in eccentricity and $\lambda$. We recover Eq.~(16) of~\cite{Iorio:2011ay}.

\section{Conclusion}\label{sec:ccl}

We have applied the Poincaré–Lindstedt method to the conservative perturbed Kepler problem and shown how the treatment of secular terms in the perturbed Binet equation directly determines the periastron advance. At linear order in a perturbation, only its first Fourier coefficient evaluated on the Keplerian orbit is required. For polynomial perturbations this coefficient is obtained algebraically and no integration is needed, while logarithmic and inverse-power perturbations can be treated using the expressions derived in Sec.~\ref{subsec:linorder}. We have also extended the procedure to higher perturbative orders and to systems involving several perturbation parameters.

We have applied the method to several conservative problems. For the point-mass model, we recovered the local-in-time periastron advance at 4PN and performed the 3PN calculation directly in standard harmonic coordinates, without needing to remove the logarithmic terms by a coordinate transformation. We also recovered known tidal, spin, and electromagnetic contributions. In addition, this method allowed us to derive new results: we derived the leading-PN contributions of arbitrary mass- and current-type tidal multipoles and arbitrary mass-type spin-induced multipoles, together with the NNLO current-type tidal quadrupole contribution and electric-dipole contributions to NNLO. The Reissner–Nordström, Schwarzschild–de Sitter and Yukawa potential illustrate the direct application of the method when the orbital dynamics is already available in Binet form.

The present work supposes the perturbing function in the Binet equation to be independent of the phase $\phi$. The non-local-in-time terms arising in the conservative dynamics at 4PN are not covered. The extension to this more general case is left for future work. 

\section*{Acknowledgments}

I am grateful to Gilles Esposito-Farèse for motivating the writing of this work, to Luc Blanchet, Guillaume Faye 
and François Larrouturou for useful discussions and feedback on the manuscript. I thank Andrea Placidi for kindly sharing files to ease comparisons. This work was supported by the Universitat de les Illes Balears (UIB) with funds from the Programa de Foment de la Recerca i la Innovació de la UIB 2024-2026 (supported by the yearly plan of the Tourist Stay Tax ITS2023-086); the Spanish Agencia Estatal de Investigación grants PID2022-138626NB-I00, RED2024-153978-E, RED2024-153735-E, funded by MICIU/AEI/10.13039/501100011033 and the ERDF/EU; and the Comunitat Autònoma de les Illes Balears through the Conselleria d'Educació i Universitats with funds from the European Union - European Regional Development Fund (ERDF) (SINCO2022/18146 - Plataforma HiTech-IAC3-BIO).

\appendix

\section{Newtonian Lagrangian for current-tidal multipoles}\label{app:Lsigma}

In this section we derive the Lagrangian associated with the $\ell^\text{th}$ tidal current multipole at its leading PN order. We start from the full GR action given in Eq.~(2.7) of~\cite{HFB19}, we remove the mass-type moments, isolate the current multipole and consider only its leading PN order, which reads
\begin{equation}
\mathcal{L}_{\sigma^{(\ell)}} = \sum_A \left[\frac{m_A v^2_A}{2} + \frac{G m_A m_B}{2r_{AB}} + \frac{\ell\,\sigma_A^{(\ell)}}{2(\ell+1)!c^2} H_L^A H_L^A\right]\,.
\end{equation}
The leading order tidal current multipole tensor $H_L^A$ is given in Eq.~(2.6b) of~\cite{HFB19}. When substituting the PN potentials regularized at the location of body $A$, one finds
\begin{equation}
H_L^A = 4 (-)^{\ell}(2\ell-1)!! \frac{G m_B}{r^{\ell+1}}\varepsilon_{j k \langle i_\ell} \hat{n}_{L-1\rangle j} v_k\,,
\end{equation}
where the brackets $\langle L \rangle = \langle i_1 \dots i_\ell \rangle$ represent the usual symmetric trace free operator. When contracting this tensor with itself, we get the current tidal multipole invariant
\begin{equation}
\bigl(H_L^A\bigr)^2 = 16 [(2\ell-1)!!]^2\frac{G^2m_B^2}{r^{2\ell+2}}\lambda_\ell\,,
\end{equation}
where the quantity $\lambda_\ell$ is given by
\begin{equation}
\lambda_\ell = L_{i_\ell} \hat{n}_{L-1} L_{\langle i_\ell} \hat{n}_{L-1\rangle}\,.
\end{equation}
Here $L_i = (n\times v)_i$, with the useful property $L_i n_i = 0$. Notice that $(L.L) = \bigl( v^2 - (nv)^2 \bigr)$. After manipulation of STF relations, one can show that $\lambda = C_\ell \bigl( v^2 - (nv)^2 \bigr)$, where 
\begin{equation}\label{eq:Cl}
C_\ell = \frac{(\ell+1)!}{2\ell(2\ell-1)!!}\,.
\end{equation}
When combining the above results, we obtain Eq.~\eqref{eq:Lsigma} and the coefficient of proportionality is fixed by
\begin{equation}
D_\ell = \frac{16[(2\ell-1)!!]^2}{(\ell-1)!(\ell+1)} C_\ell = 8(2\ell-1)!!\,.
\end{equation}

\bibliography{RefList_EccentricTides}

\end{document}